\documentstyle[12pt]{article}
\begin{document}
\thispagestyle{empty}
\begin{center} \LARGE \tt \bf {Metric-torsion preheating: cosmic dynamo mechanism?}
\end{center}
\vspace{1.0cm}
\begin{center}
{\large By L.C. Garcia de Andrade\footnote{Departamento de
F\'{\i}sica Te\'{o}rica - IF - UERJ - Rua S\~{a}o Francisco Xavier
524, Rio de Janeiro, RJ, Maracan\~{a}, CEP:20550.
e-mail:garcia@dft.if.uerj.br}}
\end{center}
\begin{abstract}
Earlier Bassett et al [Phys Rev D 63 (2001) 023506] investigated the
amplification of large scale magnetic fields during preheating and
inflation in several different models. They argued that in the
presence of conductivity resonance effect is weakened. From a dynamo
equation in spacetimes endowed with torsion recently derived by
Garcia de Andrade [Phys Lett B 711: 143 (2012)] it is shown that a
in a universe with pure torsion in Minkowski spacetime the
cosmological magnetic field is enhanced by ohmic or non-conductivity
effect, which shows that the metric-torsion effects is worth while
of being studied. In this paper we investigated the metric-torsion
preheating perturbation, which leads to the seed cosmological
magnetic field in the universe with torsion is of the order of
$B_{seed}\sim{10^{-37}Gauss}$ which is several orders of magnitude
weaker than the decoupling value obtained from pure metric
preheating of $10^{-15}Gauss$. Despite of the weakness of the
magnetic field this seed field may seed the galactic dynamo.
\end{abstract}

Key-words: Modified gravity theories, primordial magnetic fields.

\newpage
\section{Introduction}
The excitment of BICEP2 experiment at south pole \cite{1} to observe
inflation and primordial gravitational waves, and recent paper by
Bonvin et al \cite{2} by considering that BICEP2 could also have
primordial magnetic field observed leads us to revisited Bassett et
al \cite{3} where they consider that metric preheating may induce
amplification of the magnetic fields at cosmological scales as a
cosmic dynamo. As is well-known the preheating is a phase after
inflation where particles are created. Bassett et al \cite{3}
investigated the amplification of large scale magnetic fields during
preheating and inflation in different models with different
conformal and gauge couplings. In the presence of conductivity
resonance effect is weakened. From dynamo equation in spacetimes
endowed with torsion recently derived by Garcia de Andrade \cite{4}
it is shown that in a universe with pure torsion in Minkowski
spacetime the cosmological magnetic field is enhanced by ohmic or
low conductivity effects, which shows that the metric-torsion
effects is worth while of being studied. In this paper we
investigated the metric-torsion preheating perturbation, which leads
to the seed cosmological magnetic field in the universe with torsion
is of the order of $B_{seed}\sim{10^{-37}Gauss}$ several orders of
magnitude weaker than the decoupling value obtained from pure metric
preheating of $10^{-25}Gauss$. Despite of the weakness of the
magnetic field this seed field may seed the galactic dynamo. The
paper is organised as follows: In section 2 the torsionic dynamo
equation is revisited. In section 3 we present the solution of
electromagnetic wave equation for the perturbed metric in
Riemann-Cartan geometry with torsion. Conclusions are presented in
section 4.
\newpage

\section{Dynamo equation endowed with torsion}

In this section we shall consider the torsion dynamo equation
obtained recently from a parity violating torsion lagrangean
\cite{4}. Here $\textbf{T}$ is a torsion vector. The self-induction
magnetic equation in the spacetime torsion background is \cite{5}
\begin{equation}
{\partial}_{t}\textbf{B}=
{\nabla}\times[\textbf{v}\times\textbf{B}+{\eta}[{\Delta}\textbf{B}-({\nabla}.\textbf{T})\textbf{B}]
\label{1}
\end{equation}
Since we are only interested here on torsion effects given by ohmic
losses we express the dynamo equation as
\begin{equation}
{\partial}_{t}B\sim{[{\eta}({\Delta}{B}-({\nabla}.\textbf{T}){B})]}\label{2}
\end{equation}
Considering that dimensionally the LHS of this equation is
equivalent to the magnetic B field divided by the decay time
${\tau}_{torsion}$ and that the ${\Delta}\sim{-k^{2}}$ where k is
the inverse of the coherent lenght L, one obtains
\begin{equation}
\frac{1}{\tau}\sim{{\eta}[-k^{2}+ikT]}\label{3}
\end{equation}
By taking only into account torsion ohmic losses origin, one obtains
in modulus
\begin{equation}
\frac{1}{\tau}\sim{{\eta}[kT]}\label{4}
\end{equation}
Now making use of the astronomical data
$L\sim{10^{2}pc}\sim{10^{10}cm}$, the ohmic resistivity
${\eta}\sim{10^{21}cm^{2}s^{-1}}$ and torsion given by Lammerzahl
\cite{6} estimate of Hughes-Drever experiment
$T\sim{10^{-17}cm^{-1}}$ after a trivial computation one obtains
${\tau}_{torsion}\sim{10^{11}years}$, which exceeds the age of the
universe $10^{10}years$. Thus with this data the origin of magnetic
fields from torsion ohmic losses is primordial. Nevertheless if one
uses stronger ohmic losses such as $10^{26}cm^{2}s^{-1}$ this
situation changes and the time decay of magnetic fields is
$10^{9}years$ which does not exceed the universe age and therefore
an efficient dynamo mechanism is necessary for the remnant magnetic
fields present in the actual universe. Now let us compute the
cosmological magnetic field obtained from the solution of the above
dynamo equation. The solution of this equation is then given by
\begin{equation}
B_{G}\sim{{\eta}kT{\tau}_{torsion}B_{seed}}\label{5}
\end{equation}
By taking into account that the galactic magnetic field
$B_{G}\sim{10^{-6}Gauss}$ one obtains a seed field of
$B_{seed}\sim{10^{-17}Gauss}$. Now let us discuss the mild galactic
dynamos which gives rise to $10^{3}s^{-1}$ amplification of magnetic
fields and obtain stringent limits for Lorentz violation than the
one obtained so far by Kostelecky et al \cite{7}. Earlier and
Berazhiani and Dolgov \cite{8} have obtained mild galactic dynamo
with amplification factor of $10^{4}s^{-1}$, which shows that the
dynamo considered here is still more mild. To obtains this result we
start by solving the dynamo equation above to find a general
solution
\begin{equation}
{\partial}_{t}\textbf{B}(t)=
{\nabla}{\times}\textbf{B}+\frac{1}{\kappa}{\nabla}{\times}\textbf{J}-{\eta}({\nabla}.\textbf{T})\textbf{B}\label{6}
\end{equation}
where ${\eta}=\frac{1}{\kappa}$ is the ohmic loss and $\kappa$ is
the conductivity. Here $\textbf{J}={\nabla}\times{\textbf{B}}$ is
the electric current. Formally this equation can be written as
\begin{equation}
{\partial}_{t}lnB(t)= \frac{1}{L}v+\frac{\eta}{L^{2}}[1-LT]\label{7}
\end{equation}
and its solution is given by the integral
\begin{equation}
B(t)=
\int{B_{seed}e^{\frac{v}{L}+\frac{\eta}{L^{2}}[1-LT]}dt}\label{8}
\end{equation}
the exponential is considered as the protogalactic dynamo. Now let
us use the expression
\begin{equation}
B(t)= B_{seed}{\eta}\frac{T}{L}\label{9}
\end{equation}
to obtain the amplification factor ${\gamma}$ as
\begin{equation}
{\gamma}= {\eta}\frac{T}{L}\label{10}
\end{equation}
Thus to an amplification factor of $10^{4}s^{-1}$ one obtains
$T\sim{10^{-23}GeV}$, while for a mild galactic dynamo of
${\gamma}\sim{10^{3}s^{-1}}$ one obtains. Let us now compute the
magnetic field seeds $B_{seed}$ based on the last computation if the
torsion field is $T\sim{10^{-9}cm^{-1}}$ and the $L\sim{10
Mpc}\sim{10^{15}cm}$, then the magnetic field seed is
$B_{seed}\sim{10^{-10}Gauss}$. This is well within the limits for
galactic dynamo seeds.
\newpage
\section{Metric-torsion preheating as cosmic dynamos?}
In this section we discuss the magnetic field amplification due to
large metric-torsion perturbations. Here we extend the analysis of
metric perturbations due to Maroto \cite{9} to spacetimes with
torsion. Let us start with the line element for a flat FLRW model
with scalar metric perturbations in the conformal Newtonian or
longitudinal gauge
\begin{equation}
ds^{2}=
a^{2}(\eta)[-(1+2{\Phi})d{\eta}^{2}+(1-2{\Phi}){\delta}_{ij}dx^{i}dx^{j}]\label{11}
\end{equation}
where $\eta$ is the conformal time. Considering the minimal coupling
between the e.m fields tensor and torsion tensor
${S^{\alpha}}_{\mu\nu}$ as given by
$F_{\mu\nu}=2{\partial}_{[\mu}A_{\nu]}+2{S^{\alpha}}_{{\mu}{\nu}}A^{\alpha}$.
From Lagrangean for the two-field model the e.m field equation
yields
\begin{equation}
{\nabla}_{\mu}F^{\mu\nu}= 0\label{12}
\end{equation}
where ${\nabla}$ is the Riemann-Cartan connection, taking only
first-order torsion terms one obtains the following wave equation
\begin{equation}
\frac{\partial}{{\partial}\eta}[(1-2{\Phi})[{\partial}_{i}A_{0}-{\partial}_{0}A_{i}+2{S^{k}}_{i0}
A_{k}]]+\frac{{\partial}}{{\partial}x^{j}}[(1+2{\Phi})[{\partial}_{j}A_{i}-{\partial}_{i}A_{j}+
2{S^{k}}_{ji}A_{k}]] =0\label{13}
\end{equation}
where we use the relations $\sqrt{-g}=a^{4}(1-2{\Phi})$,
$g^{00}=-a^{-2}(1-2{\Phi})$,$g^{ii}=a^{-2}(1+2{\Phi})$ in the
perturbed metric (\ref{11}). From Coulomb gauge $A_{0}=0$,
${\partial}^{i}A_{i}=0$ one obtains
\begin{equation}
\frac{d^{2}}{d\eta^{2}}{A_{i}}-{\nabla}^{2}A_{i}+2(1+2{\Phi})({\partial}_{j}{S^{k}}_{ji})A_{k}+
2(1+2{\Phi}){S^{k}}_{ji}{\partial}_{j}A_{k}+\frac{{\partial}{\Phi}}{{\partial}x^{j}}{S^{k}}_{ji}A_{k}+
\frac{{\partial}{\Phi}}{{\partial}x^{j}}
{S^{k}}_{ji}{\partial}_{j}A_{k}=0 \label{14}
\end{equation}
Now by Fourier transforming this equation one obtains
\begin{equation}
\frac{d^{2}}{d\eta^{2}}{A_{i}}+k^{2}A_{k}-2\frac{d{\Phi}}{d{\eta}}\frac{dA_{k}}{d\eta}+4k^{2}{\Phi}A_{k}=
+2ik_{j}{S^{k}}_{ji}A_{k}+2(1-2{\Phi})\frac{d{S^{l}}_{k0}}{d\eta}
\label{15}
\end{equation}
splitting the real and imaginary parts one obtains the following
constraint on torsion ${S^{k}}_{ji}k_{j}=0$ and the wave equation
\begin{equation}
\frac{d^{2}}{d\eta^{2}}{A_{i}}+k^{2}A_{k}-2\frac{d{\Phi}}{d{\eta}}\frac{dA_{k}}{d\eta}+4k^{2}{\Phi}A_{k}=
+2(1-2{\Phi})\frac{d{S^{l}}_{k0}}{d\eta}A_{k}\label{16}
\end{equation}
The term $\frac{dA_{k}}{d\eta}$ can be absorbed in a electric
current term to take into account the effects of conductivity as in
Maroto \cite{9} using torsionless preheating metric perturbations.
Taking the torsion tensor as ${S^{l}}_{k0}:=
{{\delta}^{l}}_{k}S_{0}$ one reduces the last equation to
\begin{equation}
\frac{d^{2}}{d\eta^{2}}{A_{k}}+k^{2}A_{k}-2\frac{d{\Phi}}{d{\eta}}\frac{dA_{k}}{d\eta}+
4k^{2}{\Phi}A_{k}=
+2(1-2{\Phi})\frac{d{S}_{0}}{d\eta}A_{k}\label{17}
\end{equation}
Now taking the transformation \cite{3} ${\tilde{A}}_{k}:=
(1-2{\Phi})A_{k}$ one obtains from the last equation the following
solution
\begin{equation}
{\tilde{A}}_{k}(\eta)={{\tilde{A^{in}}}_{k}+\frac{1}{k}\int{[\frac{d^{2}{\Phi}}{d{\eta}^{2}}+
4\frac{d{\Phi}}{d\eta}}(1+{\Phi})S_{0}]A_{k}{\times}sin({\eta}-{\eta}^{*})d{\eta}^{*}}\label{18}
\end{equation}
This last equation shows that a proper torsion gauge can lead to a
differential equation constraint to torsion and metric scalar
perturbation $\Phi$ given by
\begin{equation}
\frac{d^{2}}{d\eta^{2}}{\Phi}+2\frac{d{\Phi}}{d{\eta}}+{\Phi}{S_{0}}=0\label{19}
\end{equation}
The solution of this equation can be simplify with the approximation
$\frac{d^{2}}{d\eta^{2}}{\Phi}\sim{0}$ yields
\begin{equation}
{\Phi}(\eta)={\Phi}_{0}e^{-S_{0}\eta} \label{20}
\end{equation}
This equation shows that for a positive torsion the metric
perturbation decays while a negative torsion induces an
amplification in the metric perturbation. Now let us relax this
strong constrain and consider the solution above in general form to
obtain the magnetic field. This can be done in analogy to Bassett et
al \cite{3}
\begin{equation}
|B_{k}|\approx{\frac{k}{a^{2}}\int{{(\frac{d{\Psi}}{d\eta})^{2}}d{\eta}}}
\label{21}
\end{equation}
where now by analogy ${\Psi}$ is given by
\begin{equation}
\frac{d^{2}{\Psi}}{d{\eta}^{2}}=[\frac{d^{2}}{d\eta^{2}}{\Phi}+4(1+{\Phi})\frac{d{\Phi}}{d{\eta}}+
{\Phi}{S_{0}}] \label{22}
\end{equation}
Thus from Bogoliubov coefficient ${\beta}_{k}$ one considers the
magnetic energy density
\begin{equation}
{\rho}_{B}={(\frac{k}{a^{4}})}^{4}{|{\beta}_{k}|^{2}} \label{23}
\end{equation}
Therefore from these last expressions one obtains finally the
magnetic field expression from solely torsion contributions
\begin{equation}
|B_{k}|\approx{\frac{kS_{0}}{a^{2}}\int{{(\frac{d{\Phi}^{2}}{d\eta})}d{\eta}}}
\label{24}
\end{equation}
which in turn yields
\begin{equation}
|B_{k}|_{torsion}\approx{\frac{kS_{0}}{a^{2}}({\Phi})^{2}}\approx{10^{-17}\frac{k}{a^{2}}
({\Phi})^{2}} \label{25}
\end{equation}
where we have used the Hughes-Drever experiment torsion result of
$S_{0}\approx{10^{-17}cm^{-1}}$. For appropriated coherent scales
such as the ones used in Bassett et al \cite{3} one obtains
$|B_{k}|_{torsion}\approx{10^{-37}G}$ which is much weaker than the
general relativistic preheating metric scalar perturbation obtained
by Bassett et al \cite{3} which is $|B_{k}|\sim{10^{-25}G}$. In
spite of that this value is still able to seed galactic dynamo.
\section{Conclusions}
The main result of this note is that primordial magnetic fields
could be of the torsion origin depending of the amount of the ohmic
losses obtained from spacetime torsion. The cosmological magnetic
field seeds range between $10^{-17}G\le{B}\le{10^{-10}G}$ which is a
more stringent limit than the ones obtained by Barrow et al
\cite{10}. An important feature of this paper is that galactic
dynamo mechanisms allowed us to place limits on Lorentz violation
similarly to what has been recently proposed by the author
\cite{11}. Actually the limit $B\sim{10^{-22}Gauss}$ which has been
recently obtained by us \cite{11}, falls very well within the
magnetic field to seed galactic dynamos as shown recently by Barrow
et al \cite{10}. It falls shortly below the interval for the B-field
given by K. Pandey et al \cite{12} for primordial magnetic field
lines using $Ly{\alpha}$ clouds. Recently R. Banerjee \cite{13} has
informed us that values of $B-field$ as low as $10^{-32}Gauss$ may
also seed galactic dynamos. Another important source of magnetic
field amplification can be metric preheating \cite{9} in a spacetime
with torsion which we discuss here in full detail. In this paper we
also show that torsion Ohmic effects also serve to amplify magnetic
fields during preheating, though the effects seems to be minimal due
to the fact that the magnetic field at decoupling is
$B_{dec}\sim{10^{-25}G}$ while respective torsion effects are
$B_{torsion}\sim{10^{-37}G}$. However, despite of the weakness of
this field this value is still able to seed galactic dynamo.
\section{Acknowledgements}
We would like to express my gratitude to D Sokoloff and A
Brandenburg for helpful discussions on the problem of dynamos and
torsion. An anonymous referee is also greatful acknowledged for
pointing out a mistake on a previous draft of this paper. Special
thanks go to my wife Telma Mussel Rabello and Dr Carlos de Souza
Lima for technical support. Financial support from CNPq. and
University of State of Rio de Janeiro (UERJ) are grateful
acknowledged.

\end{document}